\documentclass{aa}  
\usepackage[breaklinks,colorlinks,citecolor=blue]{hyperref}
\usepackage{graphicx}
\usepackage[normalem]{ulem}
\usepackage{mathrsfs,amssymb,amsmath,bm}
\usepackage[usenames]{color}
\usepackage{booktabs}
\usepackage{multirow}
\usepackage{soul}
\usepackage{natbib}

\newcommand{\be}{\begin{equation}}
\newcommand{\ee}{\end{equation}}

\begin{document} 

\title{Magnetic field dragging in filamentary molecular clouds}

\authorrunning{D. Tapinassi et al.}
\titlerunning{Magnetic field in filaments}

\author{Domitilla Tapinassi\inst{1}
\and
Daniele Galli\inst{2}
\and
Marco Padovani\inst{2}
\and
Henrik Beuther\inst{3}
}

\institute{$^1$Dipartimento di Fisica e Astronomia, Universit\`a degli Studi di Firenze, Italy \\
\email{domitilla.tapinassi@edu.unifi.it} \\
$^2$INAF--Osservatorio Astrofisico di Arcetri, Largo E. Fermi 5, 50125 Firenze, Italy \\
$^3$Max Planck Institute for Astronomy, K\"onigstuhl 17, D-69117 Heidelberg, Germany
}

\date{}

\abstract
{Maps of polarized dust emission of molecular clouds reveal the
morphology of the magnetic field associated with star-forming regions.
In particular, polarization maps of hub-filament systems show the
distortion of  magnetic field lines induced by gas flows onto and
inside filaments.}
{We aim to understand the relation between the curvature of magnetic
field lines associated with filaments in hub-filament systems and the
properties of the underlying gas flows.}
{We consider steady-state models of gas with finite electrical
resistivity flowing across a transverse magnetic field. We derive
the relation between the bending of the field lines and the flow
parameters represented by the Alfv\'en Mach number and the magnetic
Reynolds number.}
{We find that, on the scale of the filaments, the relevant parameter
for a gas of finite electrical resistivity is the magnetic Reynolds
number, and we derive the relation between the deflection angle of
the field from the initial direction (assumed perpendicular to the
filament) and the value of the electrical resistivity, due to either
Ohmic dissipation or ambipolar diffusion.}
{Application of this model to specific observations of polarized
dust emission in filamentary clouds shows that magnetic Reynolds
numbers of a few tens are required to reproduce the data. Despite
significant uncertainties in the observations (the flow speed, the
geometry and orientation of the filament), and the idealization
of the model, the specific cases considered show that ambipolar
diffusion can provide the resistivity needed to maintain a
steady state flow across magnetic fields of significant strength
over realistic time scales.}


\maketitle 

\section{Introduction}          
\label{sec_intro}

The filamentary structure of molecular clouds has recently received
significant attention, both observationally and theoretically,
particularly in relation to the morphology of the associated magnetic
field derived from optical, near-infrared and sub-millimeter
polarization maps \citep[see, e.g.,][for a recent review]{pattle23}.
Various theoretical models have been developed to describe
self-gravitating cylindrical filaments in hydrostatic equilibrium,
partially supported by magnetic fields with a poloidal, toroidal
or helical geometry \citep{nagasawa87,fiege00a,tomisaka14,toci15a,toci15b}.
While these characteristics may be suitable for describing massive
and relatively isolated filaments, a frequently observed feature
in star-forming regions is the ``hub-filament'' structure, consisting
of a network of filaments converging to a central density concentration
(the ``hub'') hosting star and cluster formation \citep[see,
e.g.,][]{myers09}.

Velocity gradients observed along filaments \citep[see, e.g.,][]{friesen13,
kirk13,pan24} suggest the presence of accretion flows possibly
driven by the gravitational attraction of dense clumps or hubs.
This is also supported by numerical simulations
\citep{zamora-aviles17,gomez18}.  However, these velocity gradients
can also be interpreted as projection of large-scale turbulence
\citep{fernandezlopez14}. If these accretion flows exist, they would
be responsible for feeding star formation in the central hub, and
for dragging the magnetic field lines, stretching and progressively
aligning them with the filament's axis
\citep{juarez17,yuan18,wang20a,koch22}.

In some high-mass star-forming regions the overall morphology of
the hub-filament system takes the form of a spiral pattern, which
could be the consequence of a coherent large-scale rotation motion
in the parent clump \citep[][]{li17,liu18,schworer19,mookerjea23}.
When available, the direction of polarized dust emission in these
regions generally follows the spiral pattern, indicating that the
magnetic field morphology is primarily shaped by the gas dynamics.
Some examples of such regions include the high-mass star-forming
complexes Monoceros\,R2 \citep{trevino19}, G9.62+0.19 \citep{dallolio19},
G327.3 \citep{beuther20}, IRAS 18089-1732 \citep{sanhueza21}, and
Monoceros\,R2 \citep{hwang22}.

In other cases of hub-filament systems there appears to be a
progressive transition in the orientation of the magnetic field.
The field is preferentially parallel to relatively low column density
filaments (also known as ``striations''), but, as column
density increases, it becomes preferentially
perpendicular to filaments \citep[see,
e.g.,][]{chapman11,cox16,arzoumanian21,chen23}. This change occurs
at a visual extinction of $A_{\rm V} \approx 2.7$\,mag, corresponding
to a column density of $N_{\rm H} \approx 10^{21.7}$\,cm$^{-2}$
\citep{planckXXXV}.  In this scenario, the magnetic field within a
filament is stretched in the longitudinal direction by the accretion
flow, taking on a strongly pinched ``hairpin''- or ``U''-shape.
Observations of bow- or U-shaped magnetic fields have been made in
various regions, such as the massive infrared-dark cloud G035.39-00.33
\citep{liu18}, the high-mass star-forming region G327.3 \citep{beuther20},
and the massive hub-filament system SDC13 \citep{wang22}.  In all
of these regions, it seems that the magnetic field has been partially
dragged by the collapsing gas flows responsible for forming the
densest structures. This scenario of dragged magnetic fields and
accretion flows within filaments finds support in high-resolution
and high-sensitivity observations of polarized dust emission. These
observations have resolved the magnetic field structure inside
filaments at scales $<0.1$\,pc. In some cases, they have revealed
a transition from perpendicular to aligned fields occurring at
$A_{\rm V} \approx 21$\,mag ($N_{\rm H}\approx 10^{22.6}$\,cm$^{-2}$)
in the Serpens South molecular cloud \citep{pillai20} and in the
hub-filament system NGC\,6334 \citep{arzoumanian21}.

In this work we investigate the configuration of the magnetic field
in and around a filament that results from dragging and bending a
pre-existing uniform magnetic field by a prescribed accretion flow.
Our motivation is to establish the relationship between observable
geometrical characteristics of the field (as derived from dust
polarization maps) and the physical properties of the filament.
First, in Sect.\,\ref{sec_wings} we consider the case of clumpy
filaments, made by perfectly conducting localized overdensities
that move toward the central hub across the cloud's magnetic field.
Then, in Sect.\,\ref{sec_diffusion} we examine the configuration
of the magnetic field in stationary accretion-diffusion flows with
constant Ohmic or ambipolar diffusion resistivity. In
Sect.\,\ref{sec_applications} we apply the methods described in
Sect.\,\ref{sec_wings}--\ref{sec_diffusion} to two specific examples,
the hub-filament structure in the Serpens South molecular cloud and
the high-mass star-forming region G327.3, and derive the magnetic
field strength from available dust polarization data. Finally, in
Sect.\,\ref{sec_disc} we discuss the implications of this work, and
in Sect.\,\ref{sec_conc} we draw our conclusions.

\section{Alfv\'en wings}
\label{sec_wings}

\cite{gomez18} explore the shape of magnetic field lines in MHD
simulations, in which filaments are long-lived structures that
channel gas flows toward a central accreting clump. Along the
spines of these filaments, transverse magnetic field lines are
stretched by the flow and assume a U-shape. The geometrical
characteristics of this U-shape depend on the flow velocity,
density and magnetic field strength.  According to \cite{gomez18},
the curvature of the field lines depends on the Alfv\'en Mach number
$M_{\rm A}=u/v_{\rm A}$, defined as the ratio of the longitudinal
flow speed $u$ and the Alfv\'en speed $v_{\rm A}=B/\sqrt{4\pi\rho}$
determined by the strength of the magnetic field $B$ and the density
$\rho$ of the filament. Therefore, by analyzing the observed curvature
of magnetic field lines (from maps of polarized dust emission) one
could potentially determine the Alfv\'en Mach number of the flow.

The relation between the geometry of the magnetic field and flow
properties has been extensively studied in the case of localized
overdensities (clumps, or ``bullets'') of density $\rho_{\rm c}$
and size $a$ moving with speed $u$ in a medium with density
$\rho<\rho_{\rm c}$ across a transverse magnetic field \citep[see,
e.g.,][]{lyutikov06, dursi08}.  The moving clump diverts the plasma
on its sides and bends the ambient magnetic field into wedge-shaped
structures similar to Cherenkov cones, called ``Alfv\'en wings.''
These wings are characterized by a deflection angle $\theta$ in the
plane containing the flow direction and the background magnetic
field, as given by
\be
\tan\theta= M_{\rm A}
\label{wings}
\ee
\citep{drell65, neubauer80}. 
The formation of Alfv\'en wings has been investigated in the context
of satellites moving in the magnetospheric plasma of a planet
\citep{kivelson07}, planets interacting with the solar wind
\citep{baumjohann10}, and galaxy clusters moving in the magnetized
intracluster medium \citep{lyutikov06}.  Equation\,(\ref{wings})
has been validated through numerical simulations \citep[see,
e.g.,][]{linker91, dursi08}. When $M_{\rm A}$ is sufficiently large,
the wings can fold over the moving clump to form a magnetotail.

This scenario is valid if the electrical resistivity of the clump
is low enough that the time scale of magnetic diffusion across the
scale $a$ is much longer than its crossing time $a/u$, and the clump
can be considered a perfect conductor.  In this case, the ambient
magnetic field is swept up and accumulates in a strongly magnetized
boundary layer of thickness approximately equal to $a/M_{\rm A}^2$
ahead of the clump, where magnetic tension approximately balances
the ram pressure gradient \citep{gomez18}.  At this point the
magnetic tension of the stretched and compressed field acts essentially
as a hydrodynamical drag, decelerating the clump on a braking time
scale $t_{\rm br}\approx \delta a/u$, where $\delta=\rho_{\rm
c}/\rho>1$ \citep{dursi08}. This is the well-known expression for
the braking time of an aligned rotator of density $\rho_{\rm c}$
in an external medium with density $\rho$ \citep{eht60,mouschovias77},
with the Alfv\'en speed in the ``external medium'' (the boundary
layer) replaced by the clump's speed $u$ by virtue of the balance
of magnetic and ram pressure. Actually, if the clump is moving at
supersonic speed, internal shocks propagating at speed $u_{\rm
s}=u/\delta^{1/2}$ promote instabilities that disrupt the clump in
a time approximately equal to $a/u_{\rm s}=\delta^{1/2} a/u$, shorter
than $t_{\rm br}$ by a factor $\delta^{1/2}$ \citep[see,
e.g.,][]{jones94,jones95}.

As recognized by \cite{gomez18}, in order for gas to flow longitudinally
along a filament across a transverse magnetic field, some of magnetic
diffusion must be involved. Without it, a fluid element of size $a$
would only be able to travel a distance approximately equal to
$\delta a$ or $\delta^{1/2}a$ before being stopped by magnetic
tension or disrupted by internal shocks. Additionally, eq.\,(\ref{wings})
suggests that large deflection angles, greater than $45^\circ$, for
example, are a result of motions at super-Alfv\'enic speeds.  Although
the determination of $M_{\rm A}$ in filament-hub system is uncertain,
current observations based on the interpretation of sub-mm dust
polarization maps and molecular line emission indicate that,
generally, $M_{\rm A} \lesssim 1$ in filaments, and $M_{\rm A}\gtrsim
1$ only in or near the central clump/hub \citep{beltran19,
hwang22,beltran24}.  In the next section we examine the effects
of a finite electrical resistivity on the motion of gas in a filament
across a transverse magnetic field.

\section{Magnetic diffusion}
\label{sec_diffusion}

A finite electrical resistivity $\eta$ substantially modifies
the situation described in Sect.\,\ref{sec_wings}. Instead of
accumulating in front of the moving clump, the ambient magnetic
field can pass through the fluid more or less unimpeded, depending
on the value of the magnetic Reynolds number
\be
R=\frac{a u}{\eta},
\label{rm}
\ee
which represents the ratio of diffusion time scale $a^2/\eta$ to
the crossing time $a/u$.  In the presence of magnetic diffusion,
the effect of flow on the geometry of the magnetic field is no
longer described by the Alfv\'en Mach number, but by the magnetic
Reynolds number. This situation can be illustrated by the following
example: consider an electrically conducting fluid with a constant
density $\rho$ and Ohmic resistivity $\eta$, flowing with a velocity
$u$ in the $x$-direction, perpendicular to a magnetic field $B$ in
the $z$-direction (see Fig.\,\ref{fig_steady}). The fluid motion
induces an electric current in the $y$-direction given by Ohm's law
as $j=ucB/(4\pi\eta)$, where $c$ is the speed of light\footnote{Faraday
unsuccessfully attempted to measure the electric current induced
by the Thames River flowing in the Earth's magnetic field
\citep{Faraday32}.}. The Lorentz force per unit mass is then $F=
jB/(c\rho)=u B^2/(4\pi\rho\eta)$, directed opposite to the fluid
motion. Therefore, the magnetic field exerts a drag on the fluid
with a magnetic damping time $t_{\rm d}=4\pi\rho\eta/B^2 =\eta/v_{\rm
A}^2$ \citep[see, e.g.,][]{roberts67, davidson01}.  The kinetic
energy per unit volume is dissipated by Joule heating at a rate
$d(\rho u^2/2)/dt=4\pi\eta j^2/c^2=\rho u^2/t_{\rm d}$. A steady
flow can be maintained provided an external driving force is acting
on the fluid on a time scale longer than $t_{\rm d}$. Alfv\'enic
disturbances generated within the flow propagate on a scale $aS$
before being damped by magnetic diffusion, where $S=R/M_{\rm A}$
is the Lundquist number.

Inspired by \cite{gomez18}, in the following section we consider a
steady flow along a filament. This flow is driven by some unspecified
external cause such as gravity, shocks, large-scale turbulence,
etc. Our goal is to compute the curvature of a magnetic field
initially perpendicular to the flow. We assume a stationary state
and neglect the back-reaction of the field on the fluid motion. In
other words, we adopt a kinematic approximation \citep{parker63},
which is the first step in a fully magnetohydrodynamical treatment.
Unlike the situation considered in Sect.\,\ref{sec_wings}, no
magnetized boundary layer is formed in this case. The fundamental
parameter in this case is the magnetic Reynolds number, rather than
the Alfv\'en Mach number.

\subsection{Diffusion-dominated gas flows}
\label{subsec_flow}

\begin{figure}
\includegraphics[width=0.5\textwidth]{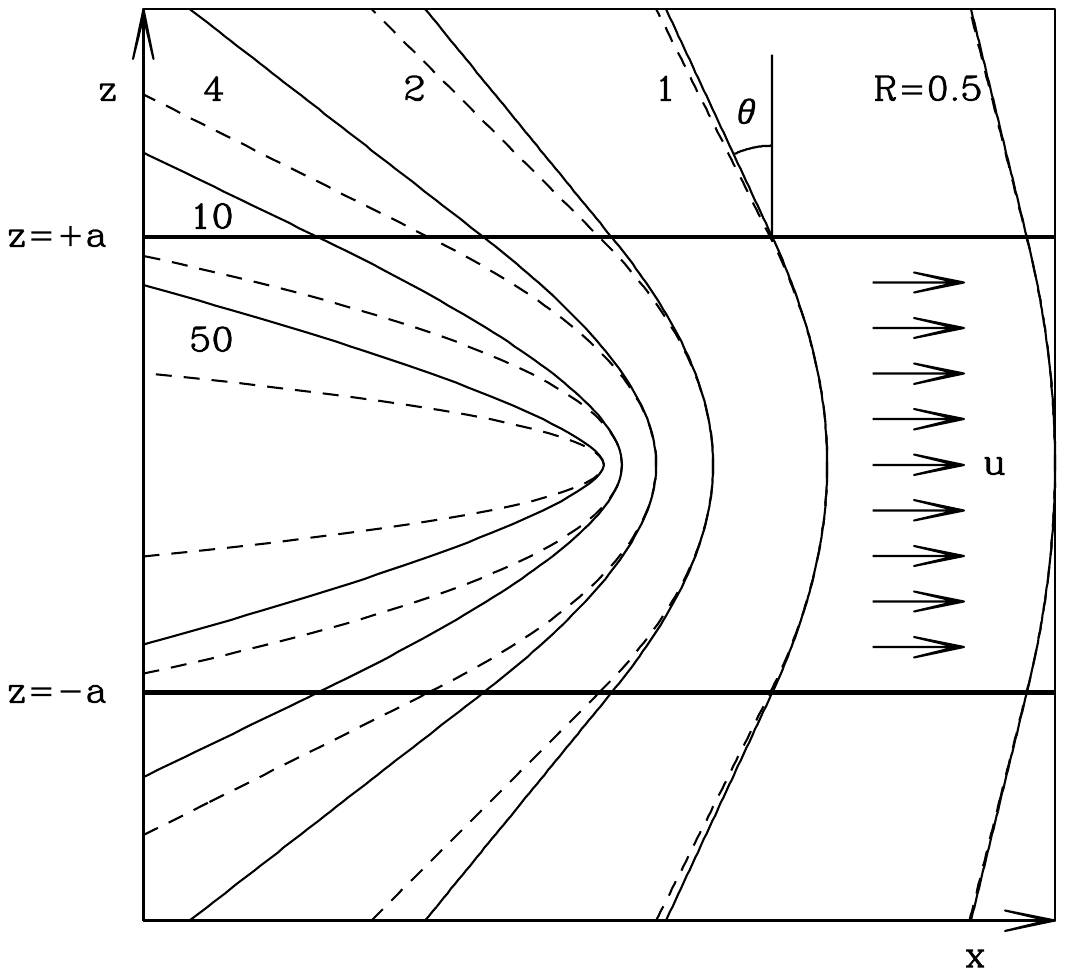}
\vspace{-3cm}
\caption{Steady-state magnetic field lines for a flow with uniform
resistivity and uniform velocity $u$ ({\it arrows}\/) in a transverse magnetic field.
for different values of the magnetic Reynolds number: $R=0.5$, 1,
2, 4, 10 and 50.  {\it Solid lines}\/: with ambipolar diffusion
($R=R_{\rm ad}$); {\it dashed lines}\/: with Ohmic resistivity
($R=R_{\rm O}$). The flow is limited to the region $-a < z < a$.
The deflection angle $\theta$ is also shown.}
\label{fig_steady}
\end{figure}

Let us consider an electrically resistive fluid flowing in the
$x$-direction with velocity $u$ between the planes $z=\pm a$, while
acting upon an initially uniform and perpendicular magnetic field
$B_0$ in the $z$-direction (see Fig.\,\ref{fig_steady}). The velocity
$u$ is assumed to vary on the scale of the filament's length, which
is much larger than the filament's width. Therefore, it can be
considered uniform over a region of size $\sim a$. Magnetic field
lines within the flow region will be stretched, progressively
increasing their curvature until they eventually reach a state where
field diffusion balances field advection, preventing any further
increase in magnetic tension.

Let us first consider the case of a uniform Ohmic resistivity
$\eta_{\rm O}$. In this case, the induction equation is linear in
the magnetic field, and can be ``uncurled''\footnote{The right-hand
side of eq.\,(\ref{ind_ohm}) should be equal to the gradient in the
$y$-direction of a function $\phi$. However, since there cannot be
a $y$-dependence in this 2-D problem, $\nabla\phi$ is at most a
constant representing a uniform electric field in the $y$-direction.
In the absence of such an external electric field, the induction
equation takes the form of eq.\,(\ref{ind_ohm}).} into
\be
{\bf u} \times {\bf B}-\eta_{\rm O}\nabla\times{\bf B}=0,
\label{ind_ohm}
\ee
where ${\bf u}=u \hat{\bf e}_x$, ${\bf B}=B_x(z) \hat{\bf e}_x +B_0
\hat{\bf e}_z$.

With the non-dimensionalization $z=a\zeta$ and $B_x(z)=B_0 b(\zeta)$
eq.~(\ref{ind_ohm}) becomes
\be
\frac{d b}{d\zeta}+R_{\rm O}=0,
\label{ind_nondim_ohm}
\ee
where $R_{\rm O}=a u/\eta_{\rm O}$ is the Ohmic magnetic Reynolds 
number. The solution of eq.~(\ref{ind_nondim_ohm}), with the 
symmetry boundary condition $b(0)=0$ is  
\be
b(\zeta)=-R_{\rm O}\zeta.
\label{ss_ohm}
\ee
As anticipated, the motion of the resistive fluid in the perpendicular
magnetic field has induced an electric current in the $y$-direction
and a magnetic field $B_x(z)=-R_{\rm O} B_0 z/a$ in the direction
of the flow, leaving unchanged the magnetic field $B_z=B_0$
perpendicular to the flow. Field lines are parabolas for $|z|<a$
and straight lines for $|z|>a$. The deflection angle of the field
lines for $|z|>a$ is equal to the magnetic Reynolds number,
\be
\tan\theta=\frac{|B_x(\pm a)|}{B_0}=|b(\pm 1)|=R_{\rm O}.
\label{theta_ohm}
\ee
The evolution of the magnetic field toward this asymptotic steady
state was computed by \cite{lundquist52}.

However, in the context of molecular clouds the magnetic Reynolds
number $R_{\rm O}$ is not relevant: Ohmic resistivity is negligible,
and the electrical resistivity is dominated by ambipolar diffusion
\citep[see, e.g.,][]{pinto08,gutierrez-vera23}.  Although ambipolar
diffusion introduces a non-linearity in the problem, a calculation
of the asymptotic configuration of the field is straightforward.
The uncurled induction equation is now
\be
{\bf u}\times {\bf B}+\frac{{\bf B}\times[{\bf B}
\times(\nabla\times{\bf B})]}{4\pi\gamma \rho_i \rho}=0,
\label{ind_ad}
\ee
where $\gamma$ is the ion-neutral drag coefficient and $\rho_i$ the
ion density. Equation\,(\ref{ind_ad}) can be non-dimensionalized as
before, obtaining
\be
(1+b^2)\frac{db}{d\zeta}+R_{\rm ad}=0,
\label{ind_nondim_ad}
\ee
where $R_{\rm ad}=au/\eta_{\rm ad}$ is the ambipolar diffusion
magnetic Reynolds number, with ambipolar diffusion
resistivity
\be
\eta_{\rm ad}=\frac{B_0^2}{4\pi\gamma\rho_i\rho},
\label{eta_ad}
\ee
assumed constant in the following.
By interchanging the dependent and independent variables,
eq.\,(\ref{ind_nondim_ad}) can be easily integrated with the boundary
condition $\zeta(b=0)=0$, resulting in
\be
b(\zeta)+\frac{1}{3}b(\zeta)^3=-R_{\rm ad}\zeta.
\ee
The solution for $b(\zeta)$ is then
\be
b(\zeta)
= \left(\frac{2}{3\alpha+\sqrt{9\alpha^2+4}}\right)^{1/3}
-\left(\frac{3\alpha+\sqrt{9\alpha^2+4}}{2}\right)^{1/3},
\ee
where $\alpha=R_{\rm ad}\zeta$.
For $|z|>a$, fieldlines are bent back 
from the direction perpendicular to the flow by an angle $\theta$ given by
\begin{align}
& \tan\theta =\frac{|B_x(\pm a)|}{B_0}=|b(\pm 1)|\nonumber \\
&=\left(\frac{3R_{\rm ad}+\sqrt{9R_{\rm ad}^2+4}}{2}\right)^{1/3}
-\left(\frac{2}{3R_{\rm ad}+\sqrt{9R_{\rm ad}^2+4}}\right)^{1/3}.
\label{theta_ad}
\end{align}

\begin{figure}
\includegraphics[width=0.5\textwidth]{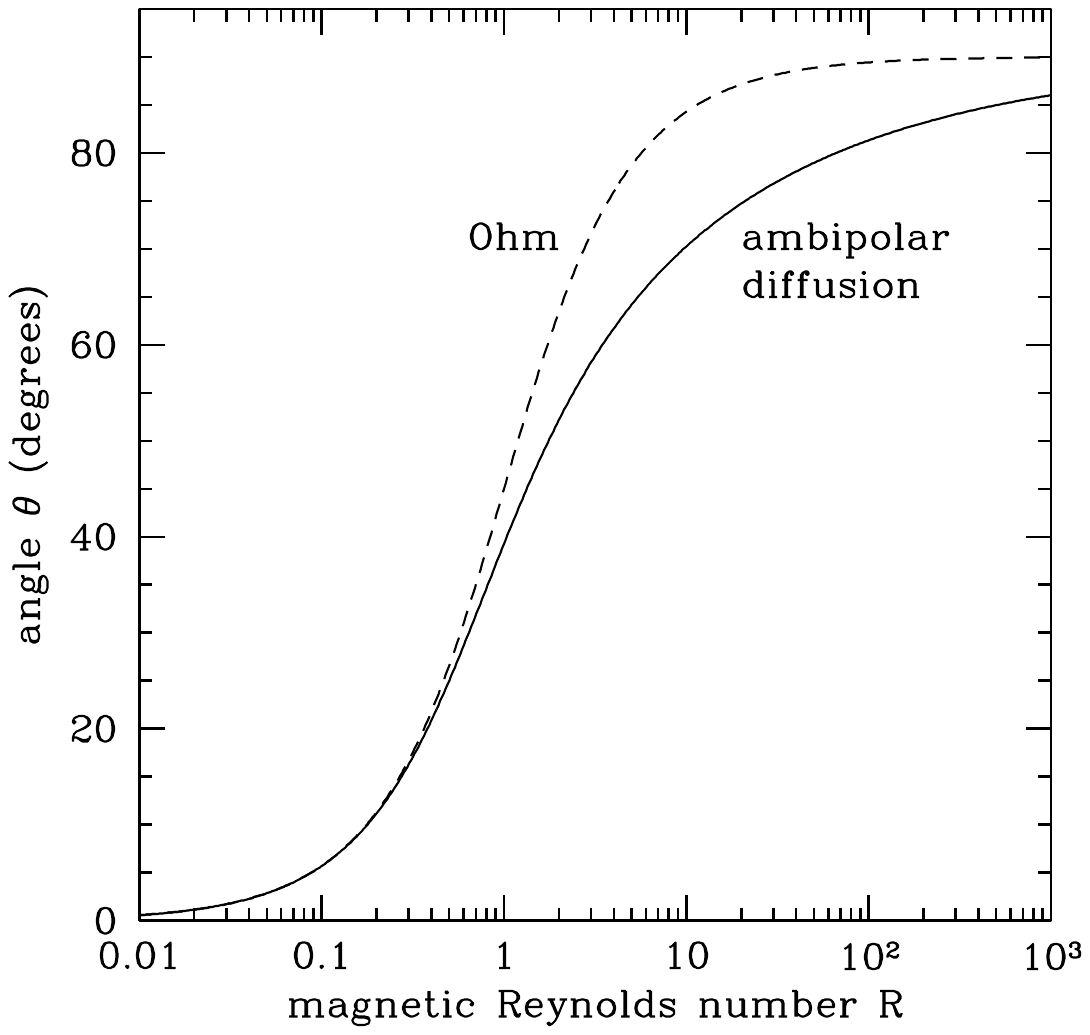}
\vspace{-3cm}
\caption{Deflection angle $\theta$ as a function of the magnetic
Reynolds number $R_{\rm O}$ (eq.\,\ref{theta_ohm}, {\it dashed line}\/) and
$R_{\rm ad}$ (eq.\,\ref{theta_ad}, {\it solid line}\/).}
\label{fig_angle}
\end{figure}

Figures\,\ref{fig_steady} and \ref{fig_angle} show the steady-state
magnetic field lines in the $x$-$z$ plane and the deflection angle
$\theta$, respectively, as a function of the magnetic Reynolds
number.  In the case with ambipolar diffusion, the flow bends the
field lines less than in the case with Ohmic resistivity for the
same magnetic Reynolds number. For small values of the magnetic
Reynolds number, the shape of the magnetic field lines is approximately
the same for both diffusive processes. In fact, for $R_{\rm ad} \ll
1$, eq.\,(\ref{theta_ad}) reduces to eq.\,(\ref{theta_ohm}),
$\tan\theta =R_{\rm ad}+{\cal O}(R_{\rm ad}^3)$.  This is also
evident from eq.\,(\ref{ind_ad}), which reduces to eq.\,(\ref{ind_ohm})
if the magnetic diffusivity is high, and therefore the induced
magnetic field is small, $b^2 \ll 1$.  For example, the condition
for the magnetocentrifugal launching of a wind from a magnetized
accretion disk, $\theta > 30^\circ$ \citep{bp82}, requires a modest
value of the radial magnetic Reynolds number: $R_{\rm O} > 1/\sqrt{3}$
from eq.\,(\ref{theta_ohm}), or $R_{\rm ad} > 10/(9\sqrt{3})$ from
eq.\,(\ref{theta_ad}).  The two values  are within 11\% of each
other.

\subsection{Numerical values}
\label{subsec_numerical}

With the usual parametrization $\rho_i=C\rho^{1/2}$, where $C$ is
a constant, the ambipolar diffusion resistivity eq.\,(\ref{eta_ad})
becomes
\be
\eta_{\rm ad}=\frac{B_0^2}{4\pi\gamma C \rho^{3/2}}.
\ee
The values of $\gamma$ and $C$ depend on the chemical composition
of the medium. However, the combination $\gamma C$ is relatively
well constrained, and can be  conveniently expressed as $\gamma
C=\chi(8\pi G)^{1/2}$, where $G$ is the gravitational constant and
$\chi \approx 1$--$3$ \citep{pinto08}. This scaling reflects
the relative importance of the ambipolar diffusion and free-fall
time scales \citep{shu83}, but in the present context in which the
self-gravity of the filament is neglected, it is just a numerically convenient
expression. 

Inserting typical numerical values, with $\rho=\mu
m_{\rm H} n$, where $\mu=2.8$ is the mean molecular weight,
\begin{align}
& \eta_{\rm ad}=6.1\times 10^{22} \chi^{-1} \nonumber \\
& \times \left(\frac{n}{10^4\,\mbox{cm$^{-3}$}}\right)^{-3/2}
\left(\frac{B_0}{\mbox{100\,$\mu$G}}\right)^{2}
\,\mbox{cm$^2$\,s$^{-1}$},
\label{ad_num}
\end{align}
and
\begin{align}
& R_{\rm ad}=\frac{au}{\eta_{\rm ad}}=0.25 \chi \nonumber \\
& \times \left(\frac{a}{0.05\,\mbox{pc}}\right)
\left(\frac{u}{\mbox{km\,s$^{-1}$}}\right)
\left(\frac{n}{10^4\,\mbox{cm$^{-3}$}}\right)^{3/2}
\left(\frac{B_0}{\mbox{100\,$\mu$G}}\right)^{-2}.
\label{reynolds_ad_num}
\end{align}
Thus, in principle, the direct evaluation of $R_{\rm ad}$ from the
curvature of the magnetic field lines allows to derive the magnetic
field strength $B_0$, provided the quantities $a$, $u$, and $n$ are
known.

Finally, the time to reach a steady state (ambipolar diffusion time) is 
\be
t_{\rm ad}=\frac{a^2}{\eta_{\rm ad}}=\frac{a R_{\rm ad}}{u},
\label{tad}
\ee
and depends, if $R_{\rm ad}$ is known, only on the filament's size $a$ and the flow velocity $u$. 

\section{Applications}
\label{sec_applications}

In this section we apply the method described in Sect.\,\ref{sec_wings}
and \ref{sec_diffusion} to two filaments in hub-filament systems,
one in the Serpens South molecular cloud and one in the G327.3
high-mass star-forming region. Each filament has a mass of a few
tens of solar masses, and is connected to a core/hub with a mass
of a few hundreds of solar masses harboring a cluster of low-mass
stars (in the Serpens South cloud) or a hot core (in G327.3). The
physical parameters of the Serpens South and G327.3 filaments derived
from observations, from the model with field-freezing
(Sect.\,\ref{sec_wings}), and from the model with ambipolar diffusion
(Sect.\,\ref{sec_diffusion}), are summarized in
Table\,\ref{tab:parameters}.

\begin{table*}
\centering
\caption{Parameters for the Serpens South and G327.3 filament: field-freezing vs. magnetic diffusion.}
\begin{tabular}{lcc|cc|ccc}
\hline
\noalign{\smallskip} \hline \noalign{\smallskip}
&  &  & \multicolumn{2}{c|}{field-freezing} & \multicolumn{3}{c}{magnetic diffusion}\\
& $u$ & $n$ &  $M_{\rm A}$ &  $t_{\rm br}^{(b)}$ & $R_{\rm ad}$ & $B_0$ &  $t_{\rm ad}$ \\
& (km\,s$^{-1})$ & ($10^5$\,cm$^{-3}$) &  & (Myr) & & (G) & (Myr) \\
\noalign{\smallskip} \hline \noalign{\smallskip}
Serpens South (FIL2) & $0.65^{+0.45}_{-0.19}$ & $1.1^{+0.90}_{-0.71}$  & $4.7^{+1.6}_{-1.0}$ & $0.75^{+0.35}_{-0.31}$ & $40^{+50}_{-20}$ & $67^{+120}_{-49}\times 10^{-6}$ & $3.0^{+6.6}_{-2.1}$\\
G327.3 (NE1--NE2) & $1^{(a)}$ & $46\pm 30$ & 3.3 & 0.18 & 16 & $1.3_{-0.70}^{+0.69}\times 10^{-3}$ & 0.28\\
       & $0.5^{(a)}$ & $46\pm 30$ & 3.3 & 0.36 & 16 & $0.95_{-0.52}^{+0.44}\times 10^{-3}$ & 0.56  \\
\noalign{\smallskip} \hline \noalign{\smallskip}
\multicolumn{8}{p{0.96\textwidth}}{\tiny {\bf Notes.} $^{(a)}$For G327.3 two values of the flow velocity have been assumed, $u=1$\,km\,s$^{-1}$ and 0.5\,km\,s$^{-1}$. $^{(b)}$A density contrast $\delta=10$ has been assumed to estimate $t_{\rm br}$ in the field-freezing case (see Sect.\,\ref{sec_wings}).} 
\end{tabular}
\vspace{1ex}
\label{tab:parameters}
\end{table*}

\subsection{Filaments in the Serpens South molecular cloud}
\label{subsec_serpens}

The Serpens South molecular cloud is a nearby star-forming region
that contains a young protostellar cluster embedded in a hub-filament
system \citep{gutermuth08}. \cite{kirk13} found that these
filaments show evidence of mass accretion flows, with rates similar
to the star formation rate in the central cluster. Specifically,
\cite{kirk13} measured a velocity gradient in the southern filament
with a value of $\nabla u_{\rm obs}=1.4\pm 0.2$\,km\,s$^{-1}$\,pc$^{-1}$,
over a projected length $L_{\rm obs}=0.33$~pc\footnote{\cite{kirk13}
assumed a distance to the Serpens cloud of 260\,pc, smaller than
the value currently adopted of 436\,pc \citep{ortiz-leon18}. However
the correction for distance cancels out in the product of $\nabla
u_{\rm obs}L_{\rm obs}$.}. This measured velocity gradient corresponds
to a flow with $u=\nabla u_{\rm obs}L_{\rm obs}/\sin\alpha$, where
$30^\circ\lesssim\alpha\lesssim 60^\circ$ is the inclination of the
filament with respect to the plane of the sky \citep{kirk13}. This
gives a flow velocity $u=0.65^{+0.45}_{-0.19}$\,km\,s$^{-1}$ (here
and in the following, the upper and lower values should be intended
as extremes of a range, not an uncertainty).

\cite{pillai20} combine near- and far-infrared observations of
polarized dust emission to map the magnetic field morphology in the
Serpens South hub-filament system. They discovered a transition in
the field orientation from approximately perpendicular to approximately
parallel to the southern filament (referred to as FIL2) at visual
extinctions above $A_{\rm V}\approx 20$\,mag up to $A_{\rm V}\approx
60$\,mag, that they interpreted as the consequence of the field
being dragged by the gas flow. Specifically, the median deviation
of polarization segments (magnetic field direction) from the parallel
direction on either side of the filament, with diameter $2a=0.1$~pc,
is $22^\circ\pm 3^\circ$, corresponding to an observed deflection
angle $\theta=78^\circ\pm 3^\circ$.

If the magnetic field pattern in the FIL2 region is interpreted in
terms of Alfv\'en wings generated by the ballistic motion of perfectly
conducting clumps in a magnetized medium, eq.\,(\ref{wings}) implies
a motion with Alfv\'en Mach number $M_{\rm A}=4.7_{-1.0}^{+1.6}$.
As mentioned in Sect.\,\ref{sec_wings}, highly super-Alfv\'enic
motions are unlikely to be present in hub-filament systems, except,
perhaps, near the central clump/hub. More importantly, as discussed
in Sect.\,\ref{sec_wings}, a frozen-in magnetic field decelerates
the fluid motion on a time scale of the order of the flow crossing
time $a/u$ times a factor $\delta$ or $\delta^{1/2}$, where $\delta$
is the density contrast between the filament and the environment.
With the values of $a$ and $u$ estimated for the FIL2 region, the
flow crossing time is $7.5^{+3.5}_{-3.1}\times 10^4$\,yr. Although
the factor $\delta$ is difficult to estimate, the deceleration time
scale of the flow in the FIL2 region appears to be quite short,
making the field-freezing model implausible.

On the other hand, in the framework of ambipolar diffusion-dominated
flows modeled in Sect.~\ref{sec_diffusion}, the observed range
of $\theta$ corresponds to $R_{\rm ad}=40^{+50}_{-20}$ (see
eq.\,\ref{theta_ad} and Fig.\,\ref{fig_angle}), an acceptable value
for dense gas \citep{mk95}.  From eq.\,(\ref{tad}), the time scale
needed to reach an advection-diffusion steady state is $t_{\rm
ad}=3.0^{+6.6}_{-2.1}$\,Myr, which is compatible with the age of
Serpens Main, the youngest cluster in the Serpens molecular cloud
\citep[$\sim 4$\,Myr,][]{zhou22}\footnote{The age of the Serpens
South cluster, the hub of the FIL2 filament, is not known.}.  In
this case, an estimate of the magnetic field strength in the FIL2
region can be obtained from eq.~(\ref{reynolds_ad_num}), if the
filament's average density $n$ is known. This quantity can be
estimated as follows. The range of visual extinction $20-60$~mag
corresponds to an observed H$_2$ column density $N_{\rm obs} =
(3.8\pm 1.9)\times 10^{22}$\,cm$^{-2}$.  Assuming a cylindrical
filament with diameter $2a$, the density $n=2N_{\rm obs}\cos\alpha/(\pi
a)$ is $n=1.1^{+0.90}_{-0.71}\times 10^5$\,cm$^{-3}$.
Eq.\,(\ref{reynolds_ad_num}) with $\chi=3$ then gives a magnetic
field strength $B_0=67^{+120}_{-49}$\,$\mu$G.  With the central
values of $N_{\rm obs}$, $\alpha$ and $B_0$, the non-dimensional
mass-to-flux ratio $\lambda\approx 7.6\,(N_{\rm
obs}\cos\alpha/10^{21}\,\mbox{cm$^{-2}$})(B_0/\mbox{$\mu$G})^{-1}$
\citep{crutcher04} is $\lambda\approx 3.0$, albeit with large
uncertainties. \cite{kusune19} apply the
Davis-Chandrasekhar-Fermi method (see Sect.\,\ref{subsec_G327}) to
model near-infrared polarization data in the same region analyzed
here, finding $B_{\rm pos}=36$\,$\mu$G and $\lambda=3.6$, compatible
with our results.

\subsection{Filaments in the high-mass star-forming region G327.3}
\label{subsec_G327}

The high-mass star-forming region G327.3 is a massive star-forming
region characterized by filamentary structures connected to a central
massive hot core. \cite{beuther20} conduct sub-mm continuum and
polarization observations of G327.3 discovering filamentary structures
characterized with U-shaped magnetic field morphologies pointing
toward the central core (see their Fig.\,1). One of these structures,
the filament NE1-NE2, has a width of $2.5^{\prime\prime}$ (equivalent
to a full size of $2a=7750$\,au at the distance of 3.1\,kpc), a
range of H$_2$ column densities $N_{\rm obs}=(4.3\pm 2.7)\times
10^{23}$\,cm$^{-2}$, and a small inclination $\alpha\approx 9^\circ$
with respect to the plane of the sky.  As in the
case of the Serpens South filament, assuming a cylindrical geometry,
the mean density is in the range $n=(4.6\pm 3.0)\times 10^6$\,cm$^{-3}$.

\begin{figure}[t]
\includegraphics[width=0.5\textwidth]{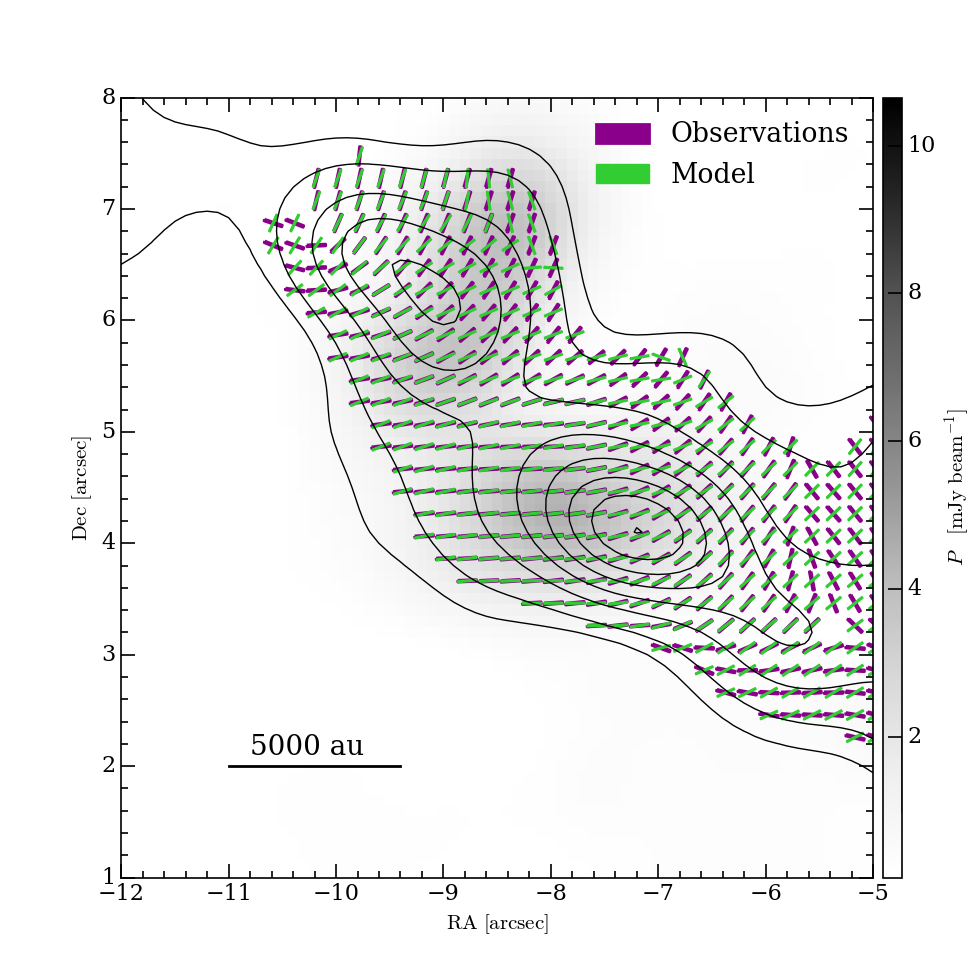}
\caption{Map of polarization segments (rotated by 90$^\circ$) and
polarized intensity $P$ ({\it greyscale}\/) in the NE1-NE2 filament
in the high-mass star-forming region G327.7: observations ({\it
purple}\/), from \cite{beuther20}; best-fit model ({\it green}\/).
The black contours show the observed 1.3\,mm dust emission intensity
starting from $3\sigma$ in steps of $6\sigma$.}
\label{fig_map}
\end{figure}

The filamentary structures in the G327.3 system were interpreted
by \cite{beuther20} as indicative of channel flows feeding star
formation in the central hot core.  To test this hypothesis with
the diffusive flow model developed in Sect.\,\ref{sec_diffusion},
we used the DustPol module of the ARTIST package \citep{padovani12}
to produce synthetic polarization maps of the filament NE1-NE2.
DustPol computes the Stokes parameters $I$, $Q$ and $U$ from the
expressions given in \cite{padovani12}, originally derived by
\cite{leedraine85}. A uniform dust temperature was assumed. The
output maps are then processed with the {\tt simobserve} and {\tt
simanalyze} tasks of the CASA
programme\footnote{\url{https://casa.nrao.edu/}}, adopting the same
antenna configuration as in the observing runs.  A best-fit of the
polarization data is obtained for a magnetic Reynolds number $R_{\rm
ad}=16$.  Figure\,\ref{fig_map} shows the comparison between the
observed polarization segments and those from the best fit model.
Unfortunately no information is available about the existence of a
longitudinal accretion flow in the filament: the almost face-on
orientation of this hub-filament system, favorable to the modeling
of dust polarization data, makes it difficult to measure a velocity
gradient. From the first moment map of $^{13}$CH$_3$CN($12_4$-$11_4$)
of \cite{beuther20}, an upper limit on the line-of-sight component
of the flow velocity $u_{\rm los} < 0.21$\,km\,s$^{-1}$ can be
derived. For an inclination $\alpha\approx 9^\circ$, this implies
$u < u_{\rm los}/\sin\alpha=1.3$\,km\,s$^{-1}$.  Assuming
$u=1$\,km\,s$^{-1}$, eq.~(\ref{reynolds_ad_num}) with $\chi=3$ gives
$B_0=1.3_{-0.70}^{+0.69}$\,mG.  With the central values of $N_{\rm
obs}$, $\alpha$ and $B_0$, the non-dimensional mass-to-flux ratio
is $\lambda\approx 2.2$. If, instead, $u=0.5$\,km\,s$^{-1}$, the
magnetic field strength would be reduced by a factor of $\sqrt{2}$,
the ambipolar diffusion time would be 2 times larger, and the
mass-to-flux ratio a factor of $\sqrt{2}$ larger
(Table\,\ref{tab:parameters}). In either case, the magnetic field
strength is higher than in the case of the Serpens South cloud (and
therefore the bending of field lines less strong), consistently
with the higher density and mass of the G327.3 star-forming region.
The time needed to reach an advection-diffusion balance, from
eq.\,(\ref{tad}), is $t_{\rm ad}=2.8$--$5.6\times 10^5$\,yr. Although
the evolutionary age of G327.3 is not known, this timescale is not
incompatible with the dynamical characteristics of the central hot
core \citep{leurini17}.

\begin{figure}[t]
\includegraphics[width=0.5\textwidth]{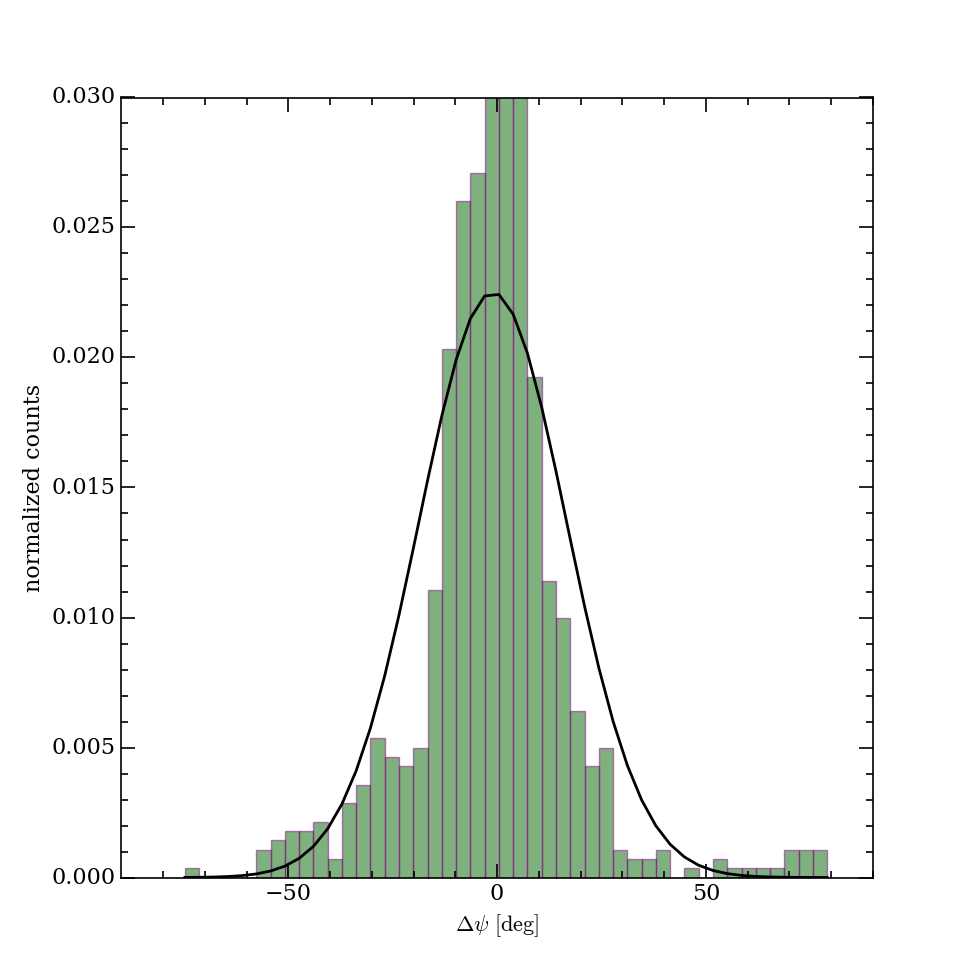}
\caption{Histogram of polarization angle residuals $\Delta\psi=\psi_{\rm
obs}-\psi_{\rm mod}$ for our best-fit model in the region shown in
Fig.\,\ref{fig_map} inside the $6\sigma$ contour of the 1.3 mm dust
continuum map. The black curve is a Gaussian with mean
$\langle\Delta\psi\rangle=-1.1^\circ$ and standard deviation
$\sigma_\psi=18^\circ$.}
\label{fig_histo}
\end{figure}

As a further check, one can estimate the strength of the component of the magnetic
field in the plane of the sky $B_{\rm pos}$ in the NE1-NE2 filament. This can be done
by applying the Davis-Chandrasekhar-Fermi formula
\citep{davis51,cf53},
\be
B_{\rm pos}=\xi\frac{\sigma_{\rm los}}{\sigma_\psi}\sqrt{4\pi\rho}.
\label{dcf}
\ee
In eq.\,(\ref{dcf}), $\xi$ is a correction factor typically set to
0.5 based on simulation of turbulent clouds \citep{ostriker01},
$\sigma_{\rm los}$ is the line-of-sight velocity dispersion, and
$\sigma_\psi$ is the standard deviation of polarization angle
residuals. Figure\,\ref{fig_histo} shows the distribution of
polarization angle residuals $ \Delta\psi=\psi_{\rm obs}-\psi_{\rm
mod}$ for the best-fit model.  By fitting a Gaussian to the
distribution of residuals, we find a mean value of
$\langle\Delta\psi\rangle=-1.1^\circ$ and a standard deviation
$\sigma_\psi=18^\circ$. Unfortunately,  the $^{13}$CH$_3$CN($12_4$-$11_4$)
data of \cite{beuther20} only have a spectral resolution of
$2.7$\,km\,s$^{-1}$, and the line width $\sigma_{\rm los}$ cannot
be determined.  For the purpose of demonstration we assume $\sigma_{\rm
los}=1$\,km\,s$^{-1}$, a reasonable value for high-mass star forming
regions \citep{beltran19,beltran24}. With this value of $\sigma_{\rm
los}$, along with the range of $n$ given above, we obtain $B_{\rm
pos}=2.2^{+1.2}_{-0.7}$\,mG, which is consistent with the estimates
of $B$ derived above from the analysis of the field curvature. The
revised Davis-Chandrasekhar-Fermi formula proposed by \cite{skalidis21},
\be
B_{\rm pos}=\frac{\sigma_{\rm los}}{\sqrt{\sigma_\psi}}
\sqrt{2\pi\rho},
\ee
gives a slightly smaller value of $B_{\rm pos}=1.8^{+0.8}_{-0.6}$\,mG.
A magnetic field strength of similar magnitude has been estimated
by \cite{beltran19,beltran24} in the outer regions of the high-mass
star forming region G31.41+031.

As in the case of the Serpens South filament, an interpretation of
the polarization pattern in terms of a frozen-in magnetic field
distorted by the motions of clumps NE1 and NE2 toward the central
hot core results in a high magnetic Alfv\'en Mach number, $M_{\rm
A}=3.3$.  More importantly, the magnetic braking time of the motion
of the clumps is within a factor $\delta$ or $\delta^{1/2}$ of the
flow crossing time $a/u = 1.8$--$3.6\times 10^4$\,yr. Therefore,
as in the case of the Serpens South filament, we conclude that the
interpretation based on the diffusive model is more realistic.

\section{Discussion}
\label{sec_disc}

The deflection angle $\theta$ of magnetic field lines analyzed in
Sect.\,\ref{sec_wings} and \ref{sec_diffusion} can be considered
as a rough measure of the magnetic field curvature $\kappa=|(\widehat{\bf
B}\cdot\nabla)\widehat{\bf B}|$, where $\widehat{\bf B}={\bf B}/|{\bf
B}|$ \citep{boozer04}. This geometrical parameter is connected to
the physical characteristics of the field and the underlying flow
\citep{yang19}.  If the velocity field has a typical lifetime shorter
than the crossing time scale $a/u$, the curvature of the field
depends on the Alfv\'en Mach number of the flow $M_{\rm A}$, as
discussed in Sect.\,\ref{sec_wings}. This is the case of turbulent
velocity fields, where the time correlation extends no farther than
$a/u$.  The scaling of the magnetic field curvature with the Alfv\'en
Mach number is in fact a characteristic of magnetohydrodynamical
turbulence \citep{yuenlazarian20}.  In the opposite limit of a flow
driven on a time scale longer than the diffusion time across the
flow region, as in the cases studied here, the diffusive 2-D models
of Sect.\,\ref{sec_diffusion}, eq.\,(\ref{theta_ohm}) or (\ref{theta_ad})
provide the relationship between $\kappa$ and $R_{\rm O}$ or $R_{\rm
ad}$, respectively.

Models of diffusive magnetic field transport have frequently been
developed in the past to describe the radial flow in accretion disks
adopting an axisymmetric thin-disk geometry. \cite{wardlekonigl90}
proposed a relationship equivalent to eq.~(\ref{theta_ohm}) between
the bending of field lines and the magnetic Reynolds number of the
radial motion in their self-similar model of the Galactic center
disk\footnote{\cite{wardlekonigl90} found that a value of the
magnetic Reynolds number (referred to as $\beta\delta$) equal to
3.4 can reproduce a set of polarization measurements in the Galactic
center region.}; \cite{lubow94} numerically confirmed eq.~(\ref{theta_ohm})
for the accretion flow in a protostellar viscous disk.  In both
these models the disk half-width $H$ is proportional to the disk
radius $\varpi$, and the magnetic Reynolds number $R=Hu/\eta$ is
constant with $\varpi$: in the model of \cite{wardlekonigl90} $u$
is constant and $\eta \propto \varpi$; in the model of \cite{lubow94}
$u\propto \varpi^{-1}$ and $\eta$ is constant.

The dependence of the magnetic Reynolds number on the spatial scale
is important, because it determines the degree of coupling of the
field to the gas and the curvature of the magnetic field lines. In
a spherical free-falling cloud with $u\propto r^{-1/2}$, the magnetic
Reynolds number $R=r u/\eta\propto r^{1/2}$ increases with $r$ (for
$\eta$ constant). Therefore fieldlines are aligned with the flow
at large radii and straight and uniform close to the central protostar
where the field decouples from the gas \citep{shu06}. Conversely,
a gravity-driven free-fall flow with uniform resistivity in a
filament with constant width has a magnetic Reynolds number
$R=au/\eta\propto r^{-1/2}$ that decreases outward. At large
distance from the center of gravitational attraction, the field is
relatively unaffected by the flow (i.e., almost uniform), but becomes
increasingly coupled to the fluid (i.e., more pinched) approaching
the center of attraction \citep[see Fig.\,16 of][]{wang20a}.
A hint of increased curvature of the fieldlines can be seen in the
bottom right corner of Fig.\,\ref{fig_map}, where the filament in
the G327.7 region merges with the central core. However, for
simplicity, in this work we have considered flows with constant
$a$, $u$ and $\eta$.

A more serious simplification of our model is the neglect of
self-gravity: Fig.\,\ref{fig_map} shows the presence in the G327.7
filament of two concentrations of dust emission intensity suggesting
the onset of fragmentation, in agreement with the magnetically
super-critical state of the filament found in Sect.\,\ref{subsec_G327}.
However, the magnetic field revealed by the dust polarization pattern
seems to be unaffected by the gravitational attraction of the two
concentrations.

Finally, it should be kept in mind that the problem of reconstructing
the morphology of the magnetic field morphology from dust polarization
maps is well-known for being degenerate \citep[see, e.g.,][]{reissl18}.
Even for the simple geometry assumed in this study, projection
effects impact the applicability of the model: a U-shaped
magnetic field line may appear more or less curved than it actually
is, depending on the viewing direction, or may not appear curved
at all \citep{gomez18}; also, the derivation of the flow velocity
along the filament from the observed velocity gradient depends on
the inclination of the filament with respect to the plane of the
sky, which is generally unknown.  The discussion of
Sect.~\ref{sec_applications} highlights the need to complement dust
polarization observations with accurate kinematic and spectroscopic
data, a task that requires, among other things, the selection of
sources with known inclinations with respect to the plane of the
sky.  Another limitation of the applicability of the diffusive model
of Sect.\,\ref{sec_diffusion} is the slow convergence of $\theta$
to $90^\circ$ for large values of the magnetic Reynolds number (see
Fig.\,\ref{fig_angle}). This slow convergence implies a significant
uncertainty in the value of the magnetic Reynolds number, even for
a relatively well-constrained value of $\theta$. However, comparing
the method with specific cases can at least provide a consistency
check on the amount of magnetic diffusivity required to maintain a
steady-state filamentary accretion, as well as a rough estimate of
the time scale needed to reach such a state.

\section{Conclusions}
\label{sec_conc}

We have elaborated the idea of \cite{gomez18}, suggesting that that
the curvature of magnetic field lines in filamentary molecular
clouds, as inferred from polarization maps, could provide insights
into the properties of accretion flows that feed star formation at
the intersection of filaments (hubs). Given the clumpy appeareance
of most filaments, at first sight it seems reasonable to consider
that the frozen-in magnetic field, being dragged by moving
inhomogeneities, could form bow wakes known as Alfv\'en wings.
These wings would be characterized by an opening angle dependent
on the Alfv\'en Mach number of the flow (see Sect.\,\ref{sec_wings}).
However, the accumulation of the swept-up magnetic field in front
of a highly conducting moving body leads to a deceleration of the
flow within a time scale comparable to the flow crossing time,
implying that filaments cannot be persistent structures. Furthermore,
in the Alfv\'en wings scenario, the observed bending of the
field would imply significantly large values of the Alfv\'en Mach
number.

If, on the other hand, filaments in hub-filament systems have long
lifetimes and transport a significant amount of mass to the central
core, a steady-state accretion flow can be established on a time
scale of the order of the flow crossing time $a/u$ times the magnetic
Reynolds number.  In this steady state, the inward advection of
magnetic field, driven by an external source, is balanced by magnetic
field diffusion (see Sect.\,\ref{sec_diffusion}). Ambipolar diffusion
can provide the necessary decoupling between the magnetic field and
the flowing matter. The two examples analyzed in this study demonstrate
that this scenario is consistent with the observed data, while an
interpretation based on field-freezing appears less plausible. It
must be stressed, however, that the interpretation of velocity
gradients in terms of gas inflow along the filaments toward the
hubs needs to be fully verified by observations.

Table\,\ref{tab:parameters} summarizes the physical characteristics
of the two regions analyzed here according to the two interpretations.
Our results support a general picture in which a star-forming core
(the hub) keeps gaining mass by accretion along filaments over most
of its lifetime, without the need to accumulate all of its mass
during a pre-stellar core phase.  Being magnetically supercritical,
the filaments studied in this work are expected to disperse by
fragmentation and collapse. In the FIL2 filament in the Serpens
South cloud \cite{friesen24} identify $\sim 5$ cores gravitationally
bound, with a mass of a few solar masses each.  However, unlike
their isolated counterparts, hub-feeding filaments are strongly
affected by protostellar feedback from the forming stellar cluster,
which eventually leads to the dispersal of the network of filaments
\citep{wang10}, and by the strong gravitational pull of the central
core, as suggested by the evidence of gas acceleration in the
proximity of the hub \citep{hacar17,zhou23,sen24}.

Further validation of the basic findings in this study can be
obtained by generating synthetic dust polarization maps using models
that incorporate more realistic filament geometry and flow properties.
Since grid-based codes have a high intrinsic numerical viscosity
$\nu_{\rm num}$ of the order of $10^{23}$\,cm$^2$\,s$^{-1}$
\citep{mckee20}, albeit dependent on resolution, and have a numerical
magnetic Prandtl number $P_{\rm num}=\nu_{\rm num}/\eta_{\rm num}
\sim 1$--2, as argued by \cite{lesaffre07}, then the numerical
resistivity of MHD simulations is quite large, of the order of the
resistivity provided by ambipolar diffusion in the typical conditions
of filaments.  Therefore it is possible that current ideal MHD
simulations of filament formation and evolution are in fact showing
the effects of diffusion-dominated gas flows outlined in this work.

\begin{acknowledgements}
We thank the referee, whose comments and suggestions resulted in
an improved final version of this paper. DG and MP acknowledge
support from INAF large grant ``The role of MAGnetic fields in
MAssive star formation'' (MAGMA).
\end{acknowledgements}

\bibliographystyle{aa}
\bibliography{bfield.bib} 

\clearpage

\end{document}